\newcommand{\qo}[1]{``#1''}                              
\newcommand{\ket}[1]{|#1\rangle}                     
\newcommand{\bra}[1]{\langle #1|}                    
\documentclass[12pt,a4paper,final]{iopart}

\usepackage{iopams}  
\usepackage{graphicx}
\usepackage[breaklinks=true,colorlinks=true,linkcolor=blue,urlcolor=blue,citecolor=blue]{hyperref}

\begin{document}

\title{Achromatic orbital angular momentum generator}

\author{Fr\'ed\'eric Bouchard$^{1}$}
\address{$^1$Department of Physics, University of Ottawa, 25 Templeton, Ottawa, Ontario, K1N 6N5 Canada}

\author{Harjaspreet Mand$^{1}$}
\address{$^1$Department of Physics, University of Ottawa, 25 Templeton, Ottawa, Ontario, K1N 6N5 Canada}

\author{Mohammad Mirhosseini$^{2}$}
\address{$^2$ Institute of Optics, University of Rochester, Rochester, New York, 14627, USA}

\author[cor1]{Ebrahim Karimi$^{1}$}
\address{$^1$Department of Physics, University of Ottawa, 25 Templeton, Ottawa, Ontario, K1N 6N5 Canada}
\eads{\mailto{ekarimi@uottawa.com}}

\author{Robert W. Boyd$^{1,2}$}
\address{$^1$Department of Physics, University of Ottawa, 25 Templeton, Ottawa, Ontario, K1N 6N5 Canada}
\address{$^2$ Institute of Optics, University of Rochester, Rochester, New York, 14627, USA}

\begin{abstract}
We describe a novel approach for generating light beams that carry orbital angular momentum (OAM) by means of total internal reflection in an isotropic medium. A continuous space-varying cylindrically symmetric reflector, in the form of \textit{two glued hollow axicons}, is used to introduce a nonuniform rotation of polarisation into a linearly polarised input beam. This device acts as a full spin-to-orbital angular momentum convertor. It functions by switching the helicity of the incoming beam's polarisation, and by conservation of total angular momentum thereby generates a well-defined value of OAM. Our device is  broadband, since the phase shift due to total internal reflection is nearly independent of wavelength. We verify the broad-band behaviour by measuring the conversion efficiency of the device for three different wavelengths corresponding to the RGB colours, red, green and blue. An average conversion efficiency of $95\%$ for these three different wavelengths is observed. This device may find applications in imaging from micro- to astronomical systems where a white vortex beam is needed.
\end{abstract}

\vspace{2pc}

\section{Introduction}
Optical orbital angular momentum (OAM), corresponding to a helical phase-front of a light beam, finds numerous applications in many research areas such as imaging systems, lithography techniques, optical tweezers, and quantum information~\cite{frankearnold:08,hell:07,paterson:01,he:95,molina:07}. Marvelously, it provides multidimensional encoding onto a single photon for classical and quantum communications, which is being used for increasing the communication channel capacity~\cite{gibson:04,boyd:11,vallone:14,mir:14,barreiro:08}. However, the methods used to generate and manipulate the OAM have been substantially restricted to a few techniques that can be listed as (i) spiral phase plates (inhomogeneous dielectric media), (ii) holographic techniques, (iii) astigmatic mode converters and (iv) spin-to-orbit coupling in inhomogeneous birefringent medium~\cite{beijersbergen:94,bazhenov:92,mir:13,allen:92,marrucci:06}. Several fundamentally different concepts were also proposed and exploited to generate OAM, such as internal conical reflection in biaxial crystals and reflection from a metallic cone reflector~\cite{berry:05,mansuripur:11}. Almost all methods are wavelength dependent, and do not yield a broad-band OAM generator. However, the existence of an achromatic OAM generator would open up new possibilities for imaging techniques such as the optical vortex coronagraph. In the optical vortex coronagraph, the background light is suppressed by means of a spiral phase plate which generates an OAM value of 2.  However, spiral phase plates are wavelength dependent and thus analyzing a polychromatic astronomical object is not feasible. An achromatic OAM generator with an OAM value of 2 would lead to analysis of \textit{white} astronomical objects~\cite{foo:05}. Recently, Zhang and Qui proposed a technique based on total internal reflection to generate achromatic radial and azimuthal polarised light beams~\cite{zhang:13}.

It has been well-established that a light beam under conditions of total internal reflection acquires an incident-angle-dependent phase change, which does not exist for light going from lower to denser medium~\cite{born}. Such behaviour triggered \textit{Augustin-Jean Fresnel} to propose and design an achromatic wave retarder in 1817 that can be used to modify the polarisation state of light; specifically he invented a broadband quarter-wave plate, now referred to as the \textit{Fresnel rhomb}~\cite{born}. The refractive index for most isotropic transparent media varies very slowly in wavelength. For instance, the refractive index of commercial N-BK7 decreases with optical wavelength at a rate of $dn/d\lambda=-0.0124$ $\mu m^{-1}$. The retardation of commercial half-wave Fresnel rhombs made of N-BK7 is $0.5089\lambda$, $0.4963\lambda$  and $0.4923\lambda$ at wavelengths of $400$ nm, $1000$ nm and $1550$ nm, respectively~\cite{thorlab}. This indeed leads to an achromatic wave-retarder that does not depend on the input beam's wavelength. There are two ways to adjust the phase retardation under conditions of total internal reflection:  (1) adjusting the beam's incident angle and  (2) implementing a sequence of reflections that overall leads to a global phase equal to the sum of each phase retardation introduced by each reflection. Therefore, one can achieve any specific phase delay between transverse electric (TE, sagittal plane polarisation) and transverse magnetic (TM, tangential plane polarisation) polarisations via a sequence of total internal reflections.
\begin{figure}[t]
\begin{center}
	\includegraphics[width=7cm]{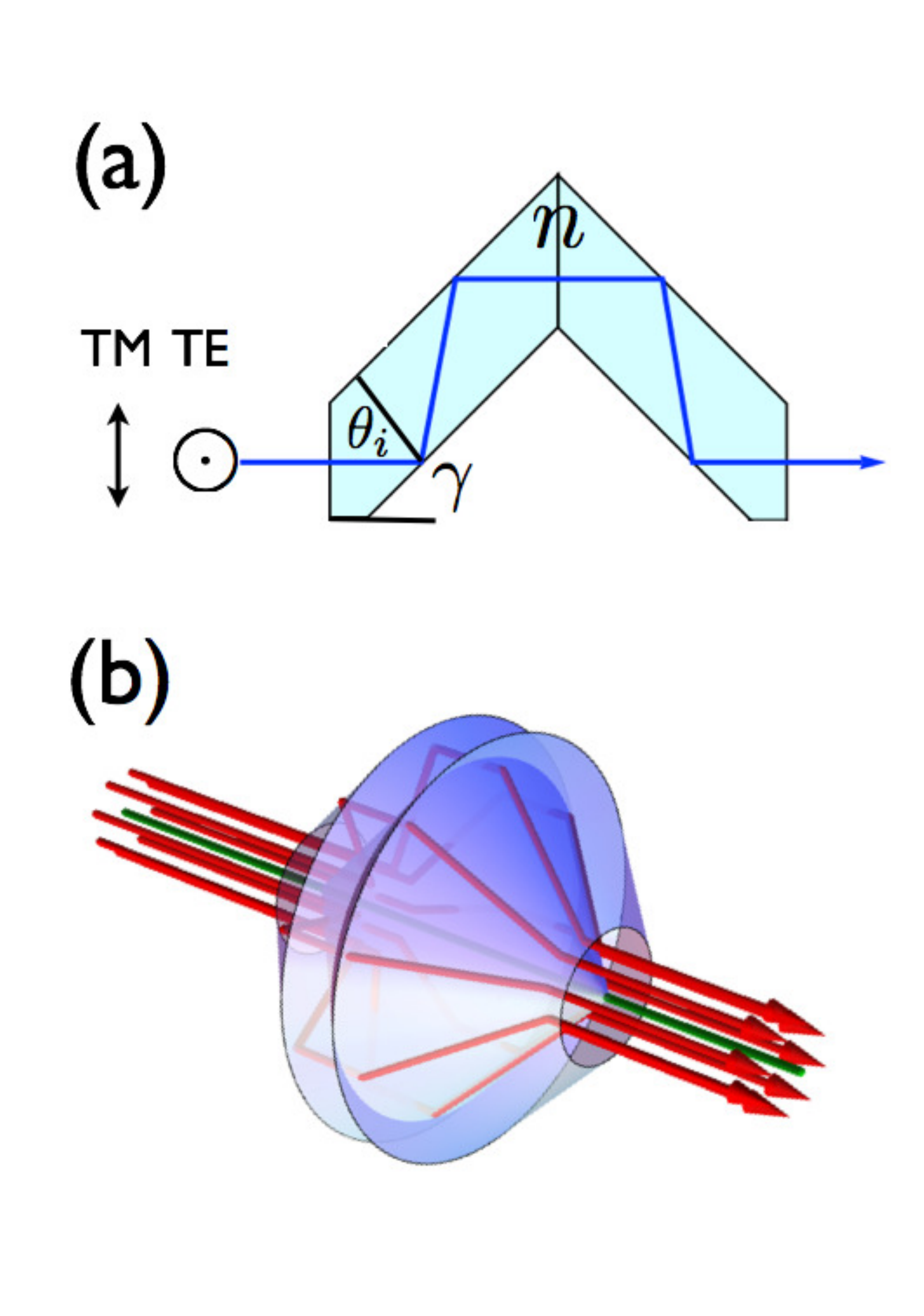}
	\caption{\label{fig:axicon}  (a) Schematic sketch of two attached Fresnel rhombs. The relative phase change between TE and TM polarisation after four appropriately designed reflections is $\pi$ radians.  (b) The three-dimensional graphical design of our device, which is a surface of revolution formed by rotating the design of part (a) about its optical axis. It is made of two glued truncated hollow axicons.}
\end{center}
\end{figure}

In this article, we propose a novel design to generate an OAM value of $\pm2$ from a circularly polarised light beam via total internal reflection in a cylindrically symmetric Fresnel rhomb-like system. A well-designed two-Fresnel-rhomb system works as a half-wave plate, which flips the helicity of input circularly polarised beam fairly independently of the beam's wavelength. Since the system is cylindrically symmetric, we would not expect any changes in the total value of angular momenta, i.e. spin+OAM. Therefore, the output light beam gains additional OAM of $\pm2\hbar$ per photon, where the sign depends on the helicity of the input beam;  $\hbar$ is the reduced Planck constant. However it is worth noticing that higher OAM values can be generated via cascading these devices with appropriate wave plates, e.g. cascading two such devices with an intermediate half-wave plate generates OAM values of $\pm4$. \\

\section{Theory}
\subsection{Total internal reflection}
Let us assume that a light beam undergoes total internal reflection at the surface of a medium having a relative refractive index of $n$ (see Figure \ref{fig:axicon}-(a)). 
The relative phase change between the two polarisation components TE and TM is given by
\begin{eqnarray}\label{eq:relative_angle}
	\delta=2 \arctan{\left(\frac{\cos{(\theta_i)}\sqrt{\sin{(\theta_i)}^2-n^2}}{\sin{(\theta_i)}^2}\right)},
\end{eqnarray}
where $\theta_i$ is the angle of incidence~\cite{born}. The beam acquires a total relative phase change between these orthogonal polarisations for a sequence of total internal reflections inside the dielectric given by the following Jones matrix:
\begin{eqnarray}\label{eq:retardation}
	{\cal J}_0= \left(\begin{array}{cc}
    1 & 0 \\ 
    0 & e^{i\delta} \\ 
  \end{array}\right).
\end{eqnarray}
In this expression, $\delta$ is the \emph{relative} phase change between two polarisation states after all reflections, which is a function of initial incidence angle, i.e. $\delta:=\delta(\theta_i,n)$.

The relative angle between the reflective surface and the incoming polarisation defines the \emph{birefringence angle} $\alpha$, which is analogous to the orientation angle of the fast axis in a wave plate. Therefore, the polarisation of the incoming beam after passing through the medium is modified by the orientation of the surface. The polarisation dependent phase shift corresponding to a rotated surface can be calculated by means of Jones calculus: ${\cal J}_\alpha=R(-\alpha)\cdot{\cal J}_0\cdot R(\alpha)$, where $R(\alpha)$ is a $2\times2$ rotation matrix~\cite{rotation}. Reflection from a flat surface implies a uniform phase variation for a set of parallel rays; instead a curved surface introduces a nonuniform phase change. This nonuniform phase alteration comes from the nonuniformity of the \emph{birefringence angle}, $\alpha$, introduced by the curved surface.  Thus, it gives rise to a nonuniform rotation of polarisation, e.g. a linearly polarised beam is transformed into a specific polarisation topology. The Jones operator of an infinitesimal reflective surface at angle $\alpha$ with respect to a fixed reference, say for example the horizontal axis, in the circular polarisation basis, is given by:
\begin{eqnarray}\label{eq:jones_general}
	{\cal J}_\alpha= \left(\begin{array}{cc}
    \cos{\left(\frac{\delta(\theta_i,n)}{2}\right)} & i\sin{\left(\frac{\delta(\theta_i,n)}{2}\right)}\,e^{-2i\alpha}\\
    i\sin{\left(\frac{\delta(\theta_i,n)}{2}\right)}\,e^{2i\alpha} & \cos{\left(\frac{\delta(\theta_i,n)}{2}\right)} \\ 
  \end{array}\right).
\end{eqnarray}
In general, $\alpha$ depends on the transverse coordinates, e.g. $\alpha:=\alpha(r,\phi)$ in polar coordinates. However, in our case we assume that the angle of the reflective surface is only a function of the azimuthal angle, i.e. $\alpha(\phi)=q\,\phi$, where $q$ is an integer or half-integer constant defined as the topological charge. A completely cylindrically symmetric reflective surface, such as that in Fig.~\ref{fig:axicon}-(b), possesses a unit topological charge, i.e. $\alpha(\phi)=\phi$. This optic, consisting of two glued hollow axicons, generates light OAM of $\ell=\pm2$, where the sign depends on the input polarisation state. Let us assume a left-handed circularly polarised input beam, i.e. $\ket{L}=(1,0)^{T}$, where $T$ stands for transpose operator. Then, the emerging beam can be split into two parts using Eq.~(\ref{eq:jones_general}), $\cos{\left(\frac{1}{2}\,{\delta(\theta_i,n)}\right)}\ket{L}+i\sin{\left(\frac{1}{2}\,{\delta(\theta_i,n)}\right)}\,e^{2i\phi}\ket{R}$, and in a similar manner it transforms a right-handed circularly polarised beam, i.e. $\ket{R}=(0,1)^{T}$, into $\cos{\left(\frac{1}{2}\,{\delta(\theta_i,n)}\right)}\ket{R}+i\sin{\left(\frac{1}{2}\,{\delta(\theta_i,n)}\right)}\,e^{-2i\phi}\ket{L}$. Thus the output polarisation, as one would expect, depends on the relative optical retardation between TE and TM polarisations.  Importantly, when it is designed to behave as a half-wave plate, i.e. $\delta=\pi$, the device does transform a left-circularly polarised beam into a right-circularly polarised beam with an OAM value of $\ell=+2$. Inversely, it transforms the right-circularly polarised beam into a left-circularly polarised beam, and gains an OAM value of $\ell=-2$. A circularly polarised light beam carries angular momentum of $\pm\hbar$  per photon. Since the device is cylindrically symmetric there should be no exchange of angular momenta with the device. In the first case, assuming a left-circularly polarised TEM$_{00}$ input beam, the total value of angular momentum is $+\hbar$ since SAM=$+\hbar$ and OAM=$0$. The device inverts the polarisation helicity into $-\hbar$, and therefore gains an OAM of $+\hbar-(-\hbar)=+2\hbar$, since total angular momentum must be conserved. Instead, for the case of a right-circularly polarised input beam, the sign changes to a negative. Indeed, the device works as a full spin-to-orbital angular momentum converter. The physics of this effect lies at the heart of a well-known quantum phenomena known as the geometric phase, where the polarisation state of photons evolves adiabatically in the Hilbert space, and it acquires an additional phase that is given by geometrical features of the Hilbert space. For our case, this geometrical phase is proportional to half of the solid angle of the evolution path in the polarisation Poincar\'e sphere~\cite{bhandari:97}. This feature has been already well-investigated in a patterned dielectric, a patterned liquid crystal cell device, and very recently in a patterned plasmonic metasurface~\cite{bosman:02,marrucci:11,karimi:14}.\newline
\begin{figure}[t]
\begin{center}
	\includegraphics[width=12cm]{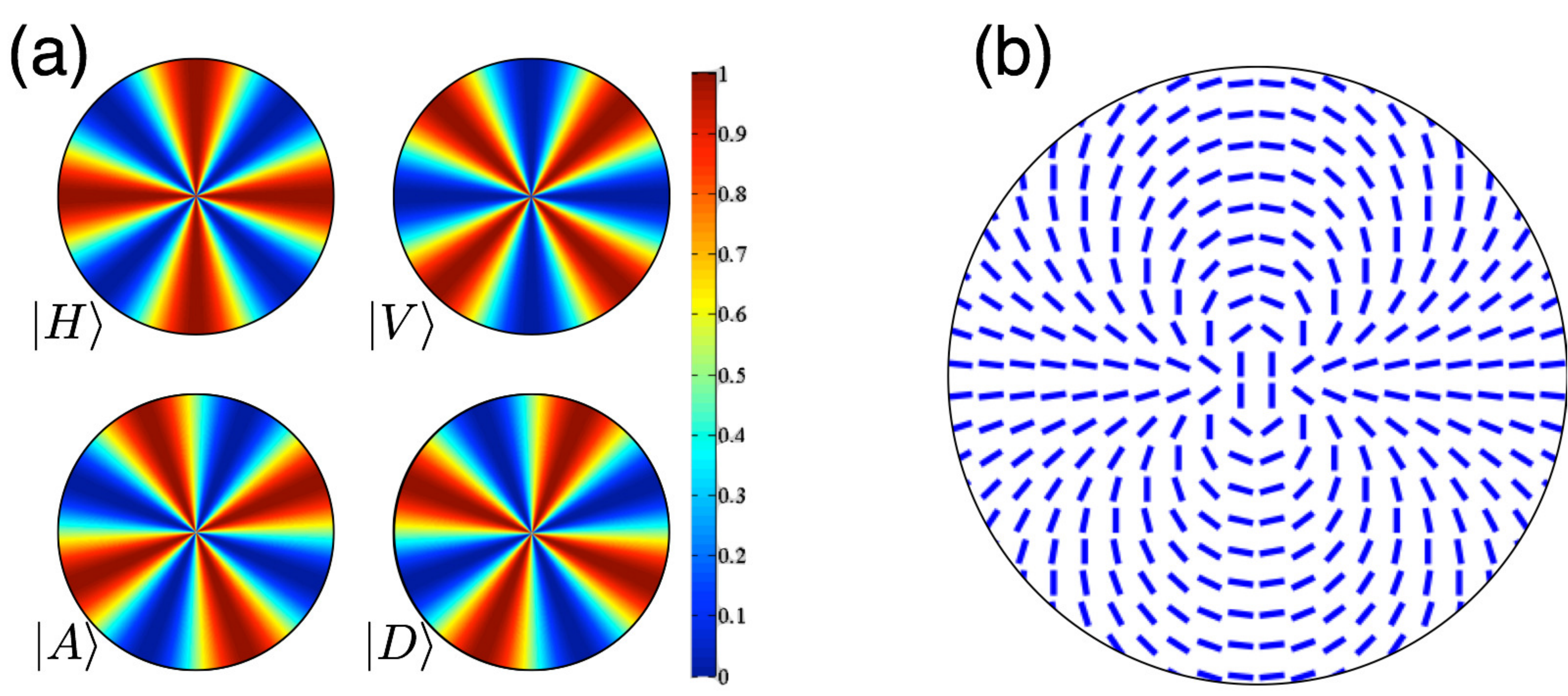}
	\caption{\label{fig:theory}  (a) Intensity profile of the generated vector beam described in section 2.2 after passing through a linear polariser oriented in Horizontal, Vertical, Diagonal and Anti-diagonal directions, respectively. (b) Transverse polarisation pattern of the generated vector beam when the device is illuminated with a horizontally polarised input beam.}
\end{center}
\end{figure}

\subsection{Vector vortex beam}
The achromatic spin-to-orbital angular momentum coupler with a specific optical retardation value of $\pi$ can also be used to generate a vector vortex beam. These types of beams have specific nonuniform transverse polarisation patterns. A subclass of these beams, cylindrical vector beams, corresponds to a superposition of two circularly polarised light beams with opposite OAM values, i.e. $\ket{L}\exp{(-i\ell\phi)}+\e^{i\gamma}\ket{R}\exp{(+i\ell\phi)}$, where $\gamma$ is a relative phase between these two beams. In the case of $|\ell|=2$, these beams can be generated by feeding our device with a linearly polarised input beam. In the circular polarisation basis, linear polarisation states can be expressed as the following superposition states: $\ket{\theta}=(e^{-i\theta}\ket{L}+e^{i\theta}\ket{R})/\sqrt{2}$, where $\theta$ is the angle of linear polarisation with respect to horizontal axis. If we apply the Jones operator ${\cal J}_\alpha$ to a specific linear polarization state, e.g. horizontal polarisation $\ket{H}:=\ket{\theta=0}$, we obtain the following output polarisation state: 
\begin{equation}
	{\cal J}_\alpha\cdot \ket{H} = \cos (2\phi) \ket{H}+\sin (2\phi) \ket{V},
\end{equation}
where a $\pi/2$ global phase is omitted. In this case we obtain a specific transverse polarisation pattern as described in Fig.~\ref{fig:theory}-(b). It is possible to reconstruct this polarisation pattern by performing polarisation state tomography, where we measure the intensity of different polarisation states using a sequence of a quarter-wave plate and a half-wave plate followed by a polarising beam splitter. Furthermore, one can determine the OAM value from these types of measurements, as it can be shown that the number \emph{intensity petals} (number of maxima) in the transverse profile is equal to twice the OAM value.
\newline

\begin{figure*}[t]
\begin{center}
	\includegraphics[width=14cm]{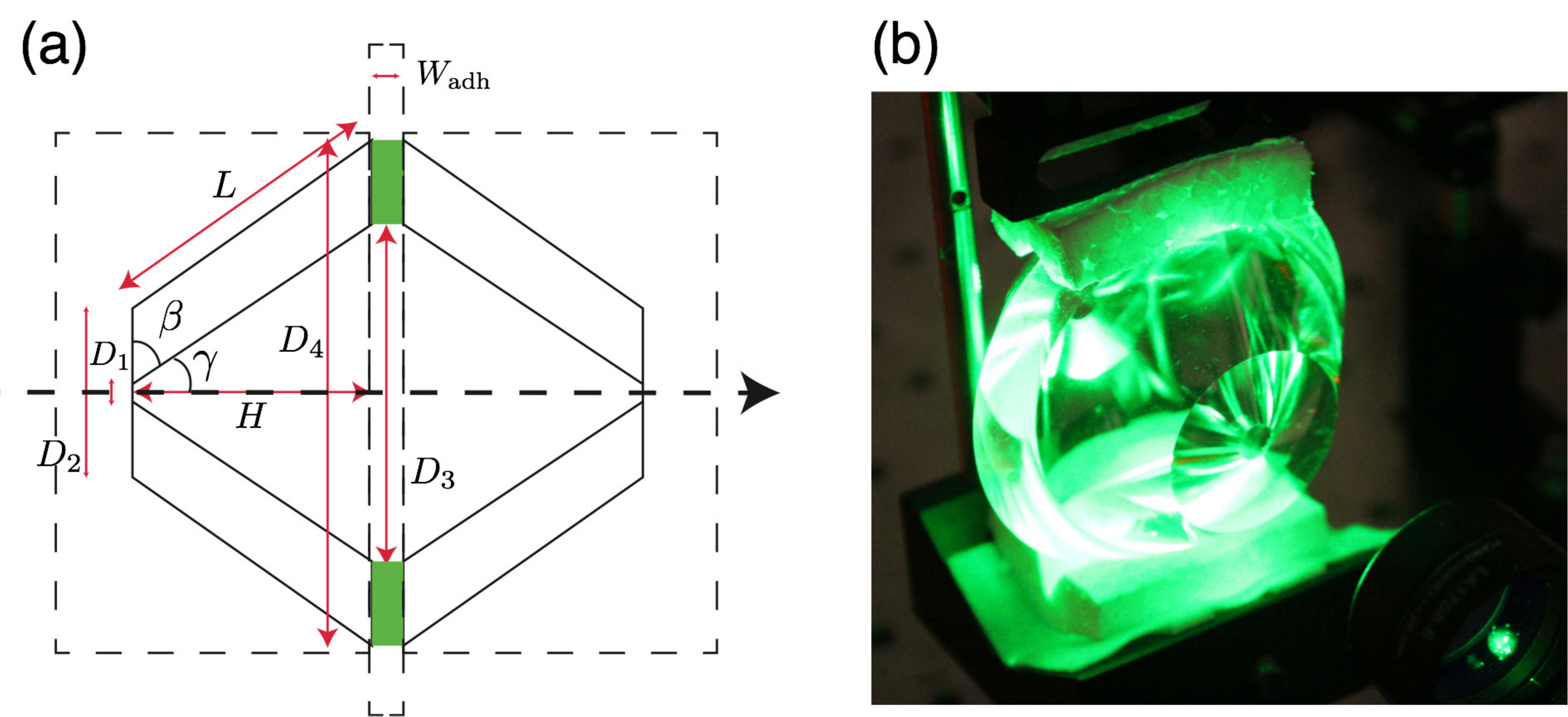}
	\caption{\label{fig:design}  (a) Two-dimensional schematic of the angular cone.  Legend: $\gamma$ is the half-angle of the cone; $\beta$ is the complementary angle with respect to $\gamma$; $W_{adh}$ is the adhesive layer thickness; $H$ and $L$ are the height and lateral height of the truncated cone; $D_1$ and $D_2$ are the inner and outer diameters of the annulus at the small end of the cone, and $D_3$ and $D_4$ are the inner and outer diameters of the annulus at the large end of the cone, respectively. (b) Photo of the manufactured device when illuminated by a green diode laser.}.
\end{center}
\end{figure*}
The retardation introduced by the reflections, i.e. $\delta(\theta_i,n)$, can be intentionally adjusted to not impose a full spin-to-orbital angular momentum conversion. Therefore, a fraction of the (circularly polarised) beam given by $\eta=\big|\sin{\left(\frac{1}{2}\delta(\theta_i,n)\right)}\big|^2$ undergoes the full spin-to-orbital angular momentum conversion, where its polarisation flips to opposite handedness and gains an OAM value of $|\ell|=2$. The remaining part, $1-\eta$, does not go through the spin-to-orbital angular momentum conversion process, and thus possesses the same polarisation and OAM value as the input beam. Implementing specific retardations, in particular having a power balance between spin-to-orbital and non spin-to-orbital angular momentum converted photons (i.e. $\eta=1/2$), leads to complex transverse polarisation patterns called \qo{\textit{polarisation-singular beams}} or \qo{\textit{Poincar\'e beams}}~\cite{cardano:13,beckley:10}. This condition requires a static design of the device, for instance by changing the dielectric material or the cone angle, and cannot be adjusted dynamically. 

Due to the presence of the \textit{surface singularity} on the axis of the device ($D_1$ in Fig.~\ref{fig:design}-(a)), a small portion of the beam near this region is scattered out. Thus, the transmitted beam possesses a doughnut shape at the exit face of the device. Nevertheless, since the emerging beam carries an OAM value of $\pm2$, this doughnut shape remains unchanged by free-space propagation. The acceptance aperture of the truncated outer cone, $D_2$ in Fig.~\ref{fig:design}-(a), introduces further diffraction on the beam, which affects the beams radial shape similar to that presented in most of OAM generators~\cite{karimi:07}. It is worth noticing that both effects are present in all OAM generator devices, except for the cylindrical mode convertor that transfers the cartesian laser cavity modes, i.e. Hermite-Gaussian, into the cylindrical solution of Laguerre-Gaussian modes~\cite{allen:92}. Furthermore, the emerging beam undergoes a type of radial astigmatism, since the rays close to the centre traverse faster than the rays at the periphery for an imperfectly collimated beam. However, this radial astigmatism can be compensated by adding a radially shaped pattern to the input and output ports of the device.\newline

\begin{table*}[h!]
\begin{center}
\begin{tabular*}{0.5\textwidth}{@{\extracolsep{\fill}} c|| c|| c }
\hline\hline
  Description&  Value & Tolerance \\
   \hline
$n$ @ 633 nm  & $1.497$ 		&  $\pm 0.001$ \\
$\beta$ 		&  $51.78^\circ$        & $\pm 0.1^\circ$   \\
$\gamma$ 	&  $38.22 ^\circ$	& $\pm 0.1^\circ$   \\
$D_1$ 		&  $6$~mm 	 	& $\pm 0.1$~mm  \\
$D_2$ 		&  $30$~mm		& $\pm 0.1$~mm  \\
$D_3$ 		&  $43.7$~mm	 	& $\pm 0.1$~mm \\
$D_4$ 		&  $67.7$~mm 		&  $\pm 0.1$~mm \\
$H$ 			&  $23.94$~mm	&  $\pm 0.1$~mm \\
$L$ 			&  $30.47$~mm	&  $\pm 0.1$~mm \\
$W_{adh}$ 	&  $\leq 25~\mu$m 	& $5~\mu$m   \\
$\Delta n$  	&  $\leq 0.1$ 	& \\
\hline\hline
\end{tabular*}
\caption{\label{tab:table1} Specifications of the device parameters. $\Delta n$ is the maximum tolerable refractive-index mismatch between the PMMA and the adhesive layer.}
\end{center}
\end{table*}
\section{Design}
The device is made of Poly methyl methacrylate (PMMA) and takes the form of two hollow cones, where the inner surface is removed by diamond turning. This approach allows us to reach a \qo{singularity diameter} as small as $D_1=6$~mm, without introducing any deleterious effects. In order to reduce any thermal birefringence, the cast acrylic was held at a constant temperature during the diamond turning process. Specifications of the fabrication of the device are given in Table~\ref{tab:table1}, where the various quantities are defined in Fig.~\ref{fig:design}. These parameters provide a clear aperture about $24$~mm. The two cones were attached to one another by a UV-curable adhesive with a thickness of $(0.025\pm0.005)$~mm, and a refractive index close to that of cast acrylic. However, some inhomogeneities were observed due to nonuniformity of the applied adhesive at the cone's interface.

\section{Results and discussions}
As described in more details below, we have experimentally demonstrated that the fabricated achromatic spin-to-orbital angular momentum coupler generates a light beam carrying an OAM value of $\ell=\pm2$. We examine the beam emerging from the device by performing polarisation projective measurements, where the nature of spin-to-orbital angular momentum coupling is used to analyze the output beam. In our analysis, the device is illuminated with different polarisation states, which are a superposition of states in the circular polarisation basis. In addition to this, we also measured the OAM spectrum of the emerging beam for different input polarisation states. This is performed by means of OAM state tomography in which the output beam was projected onto different OAM basis and then the corresponding OAM density matrix was measured. Furthermore, we show generation of a vector vortex beam where the polarisation undergoes a full rotation as we go through a full cycle around the beam's axis. The transverse polarisation pattern of the vector beam is determined by measuring the reduced Stokes parameters~\cite{cardano:12}.

\subsection{Examining the device with a laser source}
Fabrication limitations encountered in this experiment consist of a central circular orifice and of an annular region where light is not transmitted, called the \emph{dead zone}. The part of the beam that is transmitted through the central opening does not undergo the sequence of total internal reflections that lead to spin-to-orbital angular momentum coupling. This central non-converted region of the output beam can be filtered out with a sequence of a quarter-wave plate and a polarising beam splitter, since the converted and non-converted beams have orthogonal polarisations. Only the outer annular region leads to spin-to-orbital angular momentum coupling, thus generating a beam carrying orbital angular momentum or a vector vortex beam. 

Because we do not have access to the entire transverse profile of the output beam, the characteristic doughnut-shaped intensity profile is not visible. Thus, we determine the OAM value by using projective measurements on a generated vector beam with a horizontally polarised input beam. In the case of a spin-to-orbital angular momentum coupling device with a unit topological charge, we would expect to see a \emph{cross pattern} after projecting the outgoing beam onto a linear polariser. In general, the number of bright regions in the transverse intensity profile of the projective measurement, i.e. petals, is twice that of the OAM value of the beam. Rotating the polariser causes a rotation onto the four-fold petal pattern. This set of polarisation projective measurements constitutes an accurate means of determining the OAM value of an output beam in the case of a circularly polarised input beam. Furthermore, the reduced Stokes parameters, which are necessary to reconstruct the polarisation pattern of the vector vortex beam, can be calculated from this set of measurements. 
\begin{figure*}[t]
\begin{center}
	\includegraphics[width=14cm]{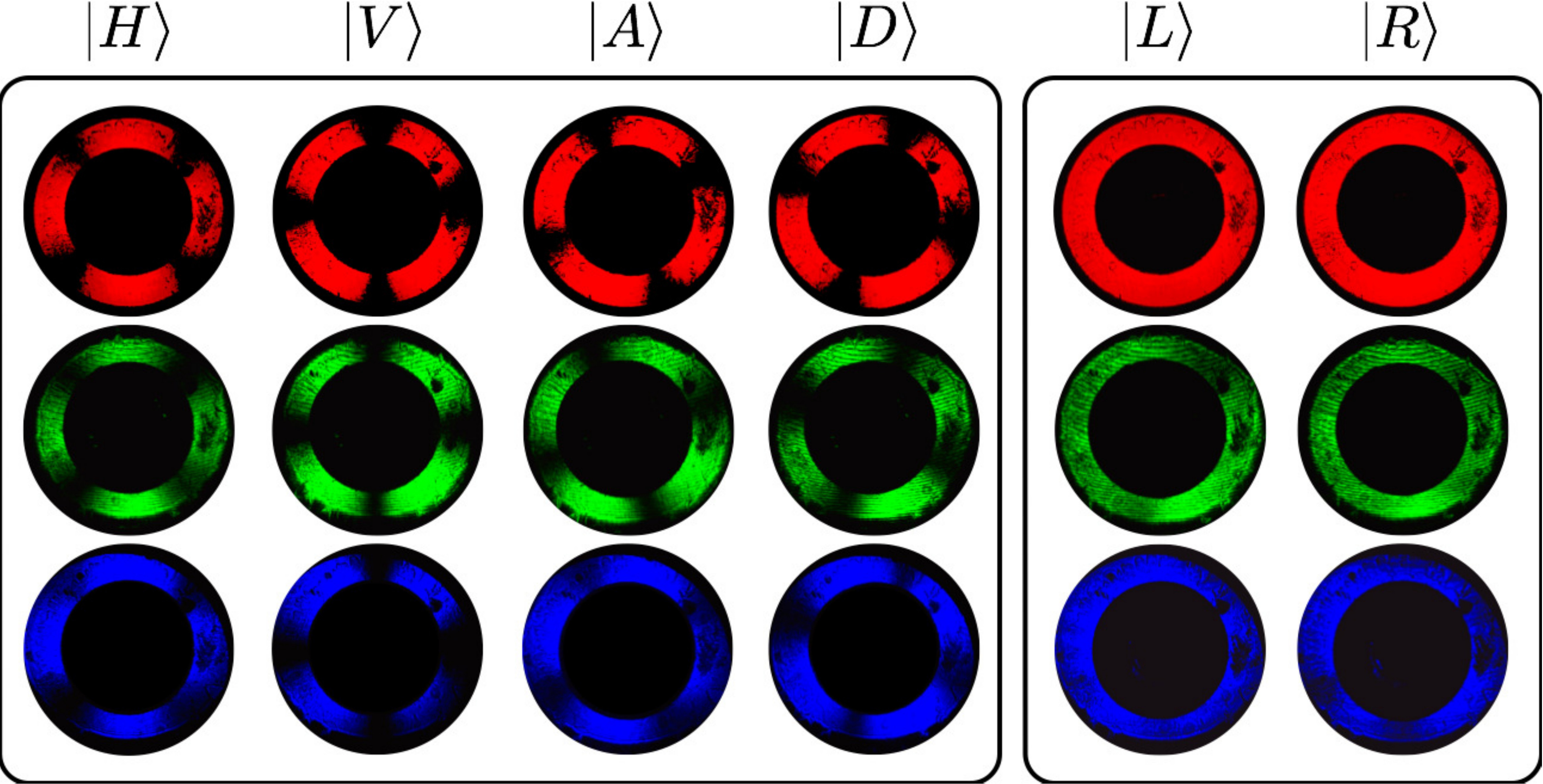}
	\caption{\label{fig:exp}  (left) Recorded intensity profiles of the outgoing beam from the device after projecting its polarisation state onto horizontal (H), vertical (V), diagonal (D), and anti-diagonal (A) directions. The device is illuminated with a linearly horizontal polarised beam. (right) Beam intensity profiles when the cone is fed with a left (L) and right (R) circularly polarised light beams. The rows correspond to different laser sources with wavelengths of $633$~nm, $532$~nm and $405$~nm, respectively. In all cases, the dark disk in the centre of the pattern is the \emph{dead zone} described in the text. In the above figure, the residual light from the central region has been artificially masked for reasons of clarity.}
\end{center}
\end{figure*}

To show the broadband performance of the device, we perform polarisation projective measurements for three different laser sources; namely red, green and blue (RGB). The red, green and blue laser sources corresponds to a He-Ne laser at a wavelength of $633$~nm and diode lasers operating at wavelengths of $532$~nm and $405$~nm, respectively. A polarising beam splitter is used to obtain a horizontally polarised input beam. Then, horizontal, vertical, diagonal and anti-diagonal components of the intensity profile are measured using a sequence of a half-wave plate and a polarising beam splitter. Figure~\ref{fig:exp} shows the desired \qo{cross} patterns  previously described. The intensity profile for left and right circularly polarised input beams are also shown in Fig.~\ref{fig:exp}, where the intensity forms a \emph{uniform} doughnut shape.  As a measure of the quality of coupling between SAM and OAM in our device, polarisation conversion efficiency (the amount of light converting from one polarisation handedness to the opposite one in circular polarisation basis) is measured. Conversion efficiencies of 96\%, 95\% and 94\% for red, green and blue light are obtained, respectively. These conversion efficiencies are quite high and indicative of the high quality of the OAM generator.

\subsection{OAM state tomography}

Polarisation projective measurements shown in the previous section give a nice illustration of the spin-to-orbital angular momentum coupling nature of the device. In this section, we show quantitative measurements confirming the generation of OAM values of $2\hbar$ per photon for a left-handed circularly polarised beam, and $-2\hbar$ per photon for a right-handed circularly polarised beam. In order to reconstruct the OAM state of the output beam we performed OAM state tomography. We measured the OAM spectrum of the output beam by performing a phase-flattening projective measurement technique~\cite{anton:01}. The measured OAM power spectrum is negligible for $|\ell|\neq 2$. The results confirm that the emerging beam from the device carries OAM values of $\ell=\pm2$. We, thus, limit our analysis to a bi-dimensional OAM Hilbert space spanned by the following basis: $\{\ket{\ell=+2}_o, \ket{\ell=-2}_o \}$, where $\ket{\ell}_o$ represents an OAM state in this Hilbert space. Hence the OAM state of the output beam can be represented by $\ket{\psi}_o=\alpha\ket{+2}_o+\beta\ket{-2}_o$, where $\alpha$ and $\beta$ are complex numbers. This specific OAM sub-space is isomorphous to the polarisation Hilbert space spanned by two orthogonal polarisation states, which can be represented by the SU(2) group~\cite{miles:99}. Pauli Matrices $\hat{\sigma}_x$, $\hat{\sigma}_y$, $\hat{\sigma}_z$, and the identity matrix $\hat{I}$ are corresponding generators of the SU(2) group. Therefore, one can reconstruct the OAM density matrix of the output light beam, i.e. $\rho_{\ket{\psi}_o}:=\ket{\psi}_o\bra{\psi}_o$, by projecting the unknown OAM state $\ket{\psi}_o$ over the eigenstates of the generators. This is identical to measuring the polarisation Stokes parameters, i.e. projecting the state on the $H$, $V$, $A$, $D$, $L$ and $R$ basis. The correspondance with the case of polarisation is straightforwardly done by associating $\ket{+2}_o$ and $\ket{-2}_o$ to the left and right-handed circular polarised states of light, respectively. A full characterization of the unknown OAM state can be done by measuring four independent \emph{Stokes-like} parameters $\Lambda_i$ ($i=0,1,2,3$) defined as:
\begin{eqnarray}\label{eq:stokes}
\left\{
\begin{array}{c}
	\Lambda_0=P_{\ket{+2}_o}+P_{\ket{-2}_o}\\
	\Lambda_1=P_{\ket{h}_o}-P_{\ket{v}_o}\\
	\Lambda_2=P_{\ket{a}_o}-P_{\ket{d}_o}\\
	\Lambda_3=P_{\ket{+2}_o}-P_{\ket{-2}_o}.
\end{array}
\right.
\end{eqnarray}
Where $P_{\ket{i}_o}$ are the power of the projective measurement over the state of ${\ket{i}_o}$, and ${\ket{h}_o}=\left({\ket{+2}_o+{\ket{-2}_o}}\right)/\sqrt{2}$, ${\ket{v}_o}=-i\left({\ket{+2}_o-{\ket{-2}_o}}\right)/\sqrt{2}$, ${\ket{a}_o}=\left({\ket{+2}_o+i{\ket{-2}_o}}\right)/\sqrt{2}$, and ${\ket{d}_o}=\left({\ket{+2}_o-i{\ket{-2}_o}}\right)/\sqrt{2}$ are the eigenstates of the $\hat{\sigma}_x$ and $\hat{\sigma}_y$, respectively~\cite{nagali:09b}. In order to achieve these set of measurements, we implement a well-known phase-flattening projective measurement. In this technique, the emerging beam from the device is imaged onto a spatial light modulator (SLM) where different computer-generated holograms corresponding to the above six measurements were displayed. These holograms \qo{flatten} the corresponding \emph{complementary phase-front} of the beam at the first order of diffraction. Thus,  only the \qo{flattened} component of the optical field can be coupled into a single mode optical fibre, which results in measuring $P_{\ket{i}_o}$~\cite{hammam:14,steve:14}. 
\begin{figure*}[t]
\begin{center}
	\includegraphics[width=15cm]{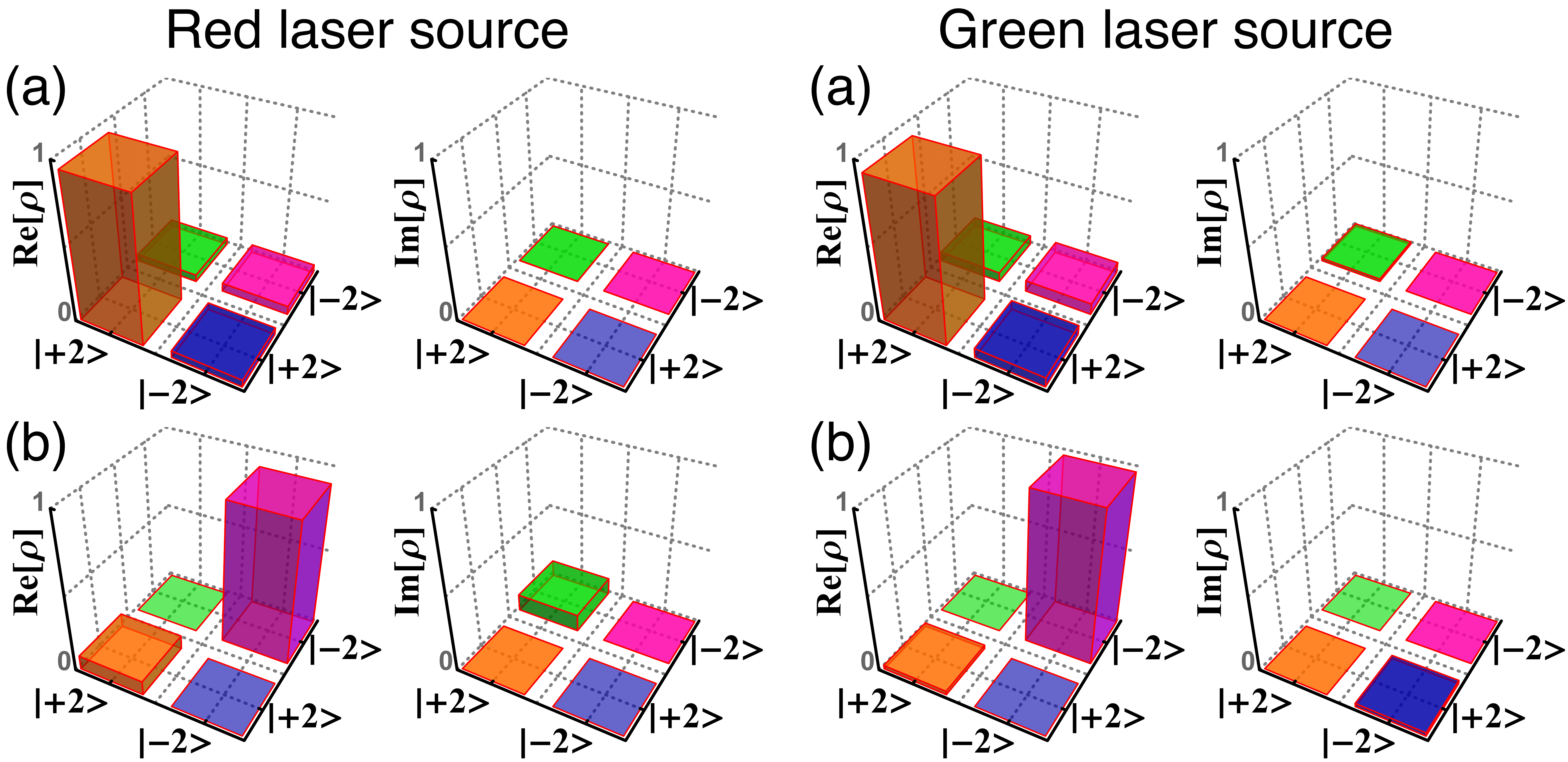}
	\caption{\label{fig:density}  Experimental OAM density matrix of the emerging beam from the device reconstructed from OAM state tomography. (a) and (b) are corresponding to left and right-handed input circular polarisations, respectively. We observe fidelity of $94\%$ and $91\%$ for generating OAM of $+2$ and $-2$ with a red laser source, and of $93\%$ and $98\%$ for generating OAM of $+2$ and $-2$ with a green laser source.}
\end{center}
\end{figure*}
The SLM and the single mode optical fibre act as the polariser in the case of polarisation state tomography. We implement this approach using a Pluto HOLOEYE SLM suitable for visible, excluding the ultra-violet, wavelengths. The power of the projective measurements $P_{\ket{i}_o}$ are recovered by a Newport power meter. From these measurements  the Stokes-like parameters of Eq.~(\ref{eq:stokes}) are calculated. The reconstructed OAM density matrix of the output beam is then calculated by the following expression:
\begin{equation}
	\hat{\rho}=\frac{1}{2} \sum_{i=0}^3 \frac{\Lambda_i}{\Lambda_0}\hat{\sigma}_i.
\end{equation}

Figure~(\ref{fig:density}) shows the reconstructed OAM density matrix for different input polarisation states, and a red and green laser source. Fidelity of $94\%$ and $91\%$ are observed for generating OAM value of $+2$ and $-2$ with a red laser source, and $93\%$ and $98\%$ are observed for generating OAM value of $+2$ and $-2$ with a green laser source. It is worth mentioning that we do not have access to a UV-SLM since all commercial liquid-crystal devices will not work as desired for the case of blue/UV light. This is based on the fact that they are originally made using UV light, and consequently, illuminating them with a UV light source would change the orientation of the polymer substrate. However, the analysis reported in the pervious section confirmed the existence of spin-to-orbital angular momentum coupling for the blue laser as well. Thus, our device can be used to generate OAM values of $\pm2$ at the UV wavelength regime, which is a missing fact in the community.

\subsection{White vector vortex beam}
We examine the broadband performance of the device by performing a polarisation projective measurement for the case of a linearly polarised white-light input beam. A liquid crystal display (laptop screen) is used as the polarised white light source and a high definition camera was used to detect the transverse intensity profile of the output beam after projecting the light polarisation state by means of a broadband half-wave plate and a polarising beam splitter. The \qo{cross} pattern was also observed for the white light source; rotating the half-wave plate leads to a rotation of the cross through the same angle.
\begin{figure}[t]
\begin{center}
	\includegraphics[width=14cm]{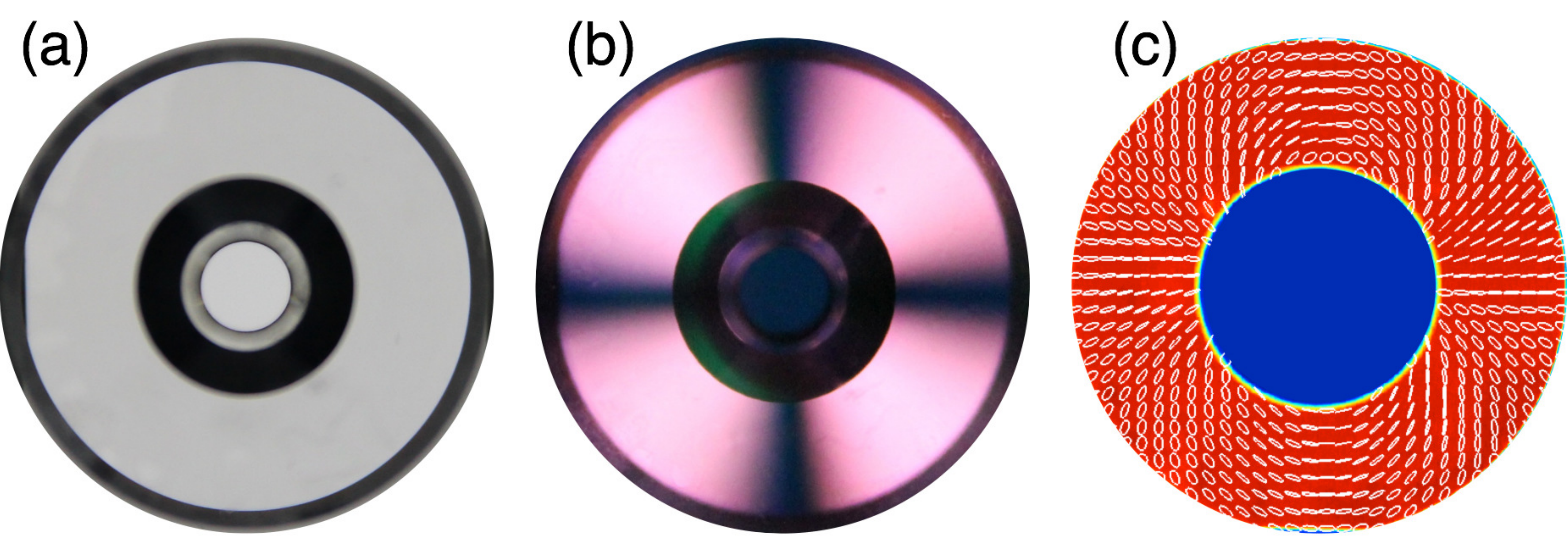}
	\caption{\label{fig:white}  (a) Intensity profile of the outgoing beam from our device when illuminated with a white polarised light source. (b) Intensity distribution of the white light beam when the the device is sandwiched between two crossed polarisers. (c) Polarisation distribution of the outgoing beam when illuminated by a white polarised light source. This beam possesses a polarisation topology of two which means that the orientation of the polarisation in the transverse plane undergoes a full rotation as we go through a full cycle around the beam's axis.}
\end{center}
\end{figure}

Figure~\ref{fig:white}-(a) shows the transverse intensity distribution of the outgoing beam from the device when illuminated with a white polarised source. As can be seen, the transmitted beam has a uniform intensity distribution. The petal pattern is also visible when the device is sandwiched between two crossed polarisers, see Fig.~\ref{fig:white}-(b). Polychromatic effects around the null intensity lines are observed, which lies at the heart of diffraction. When the device is illuminated with a polychromatic (white) wave, different colours undergo different diffraction effects since they have different wavelengths~\cite{gbur:01,amiri:05}. As we already discussed, the device transfers a linearly polarised beam into a vector vortex beam. We reconstructed the polarisation pattern of the generated vector beam by performing polarisation state tomography, where the reduced Stokes parameters of the output beam are measured via polarisation projective measurements, see Ref.~\cite{cardano:12} for more details. The reconstructed topology of the generated beam is shown in Fig.~\ref{fig:white}-(c).  As can be seen, the transverse polarisation pattern corresponds to the expected \qo{dipole} shape shown previously in the theory section. 

\section{Conclusions}
In conclusion, we have proposed and demonstrated an efficient and novel way to convert the spin of incoming light into OAM via spin-to-orbital angular momentum coupling in an \emph{isotropic} media. The idea lies at the heart of the space-variying cylindrically symmetric dielectric slab formed by two hollow axicons under a specific total internal reflection condition, which works as an inhomogeneous half-wave plate. Since the relative phase change is almost wavelength independent, the device works as an achromatic OAM generator in the OAM subspace of $|\ell|=2$ for visible and infrared regime. Moreover, the device introduces a specific polarisation topology into the incoming beam, since it possesses a space-variant polarisation dependent phase shift. Thus, as we have shown, it can be used to generate a white light vector vortex beam. This specific OAM generator can also be widely used in imaging techniques such as an optical vortex coronagraph, where achromatic OAM generators of $|\ell|=2$ is needed~\cite{foo:05}.\\

The authors would like to thank an anonymous reviewer for bringing to our attention the related work published in references ~\cite{Tosh:12} and ~\cite{Tosh:14}.  However, we note that this earlier work deals with the generation of achromatic, radially polarised light beams, whereas our work deals with the development of a device capable of generating beams with an OAM of $|\ell|=2$. The novelty of our work is therefor not compromised by this other work.

\section{Acknowledgments}
The authors thank Joe Vornehm, Mark Martucci and Doug Hand for insightful discussions regarding the cone fabrication. The authors also acknowledge the support of the Canada Excellence Research Chairs (CERC) program.


\section*{References}
\begin{thebibliography}{99}

\bibitem{frankearnold:08} 
Franke-Arnold S, Allen L and Padgett M J 2008 \textit{Laser \& Photon. Rev.} {\bf 2} 299.

\bibitem{hell:07} 
Hell S W 2007 \textit{Science} {\bf 316} 1153.

\bibitem{paterson:01} 
Paterson L, MacDonald M P, Arlt J, Sibbett W, Bryant P E and Dholakia K 2001 \textit{Science} {\bf 292} 912.

\bibitem{he:95} 
He H, Heckenberg N R and Rubinsztein-Dunlop H 1995 \textit{J. Mod. Opt.} {\bf 42} 217.

\bibitem{molina:07} 
Molina-Terriza G, Torres J P and Torner L, 2007 \textit{Nat.\ Phys.} {\bf 3} 305.

\bibitem{gibson:04} 
Gibson G, Courtial J, Padgett M J, Vasnetsov M, Pasko V, Barnett S M and Franke-Arnold S 2004 \textit{Opt. Express} {\bf 12} 5448.

\bibitem{boyd:11} 
Boyd R W, Jha A, Malik M, O'Sullivan C, Rodenburg B and Gauthier D J 2011 \textit{Proc. of SPIE} {\bf 7948} 79480L-1.

\bibitem{vallone:14}
Vallone G, D'Ambrosio V, Sponselli A, Slussarenko S, Marrucci L, Sciarrino F and Villoresi P 2014 \textit{Phys. Rev. Lett.} {\bf 113} 060503.

\bibitem{mir:14}
Mirhosseini M, Maga–a-Loaiza O S, O'Sullivan M N, Rodenburg B, Malik M, Lavery M P J, Padgett M J, Gauthier D J and Boyd R W 2014 \textit{arXiv:1402.7113}.

\bibitem{barreiro:08} 
Barreiro J T, Wei T C and Kwiat P G, 2008 \textit{Nat.\ Phys.} {\bf 4} 282.

\bibitem{beijersbergen:94} 
Beijersbergen M W, Coerwinkel R P C, Kristensen M and Woerdman J P 1994 \textit{Opt. Comm.} {\bf 112} 321.

\bibitem{bazhenov:92} 
Bazhenov V Yu, Soskin M S and Vasnetsov M V, 1992 \textit{J. Mod. Opt.} {\bf 39} 985.

\bibitem{mir:13} 
Mirhosseini M, Magana-Loaiza O S, Chen C, Rodenburg B, Malik M and Boyd R W, 2013 \textit{Opt. Express} {\bf 21} 30196.

\bibitem{allen:92} 
Allen L, Beijersbergen M W, Spreeuw R and Woerdman J P 1992 \textit{Phys. Rev. A} {\bf 45} 8185.

\bibitem{marrucci:06} 
Marrucci L, Manzo C and Paparo D, 2006 \textit{Phys. Rev. Lett.} {\bf 96} 163905.

\bibitem{berry:05} 
Berry M V, Jeffrey M R and Mansuripur M 2005 \textit{J. Opt. A: Pure Appl. Opt.} {\bf 7} 685.

\bibitem{mansuripur:11} 
Mansuripur M, Zakharian A R and Wright E M 2011 \textit{Phys. Rev. A}  {\bf 84}, 033813.

\bibitem{foo:05} 
Foo G, Palacios D M and Swartzlander G A 2005 \textit{Opt. Lett.}  {\bf 30} 3308.

\bibitem{zhang:13} 
Zhang X and Qui L 2013 \textit{Opt. Eng.} {\bf 52}, 048001.

\bibitem{born}
Born M and Wolf E \textit{Principles of Optics}, (Cambridge University Press; 7th edition 1999).

\bibitem{thorlab}
These data were taken from manufactured Fresnel rhombs (Product number of FR600HM) with N-BK7 martial form Thorlabs company.   

\bibitem{rotation}
Active $2\times2$ rotation matrix about $z$-axes is 
$R(\alpha)=\left[\begin{array}{cc}
\cos{\alpha} & \sin{\alpha} \cr 
-\sin{\alpha} & \cos{\alpha} \cr 
\end{array}\right]$.

\bibitem{bhandari:97} 
Bhandari R 1997 \textit{Phys. Rep.} {\bf 281} 1.

\bibitem{bosman:02} 
Bomzon Z, Biener G, Kleiner V and Hasman E 2002 \textit{Opt Lett.} {\bf 27} 1141.

\bibitem{marrucci:11} 
Marrucci L, Karimi E, Slussarenko S, Piccirillo B, Santamato E, Nagali E and Sciarrino F 2011 \textit{J. Opt.} {\bf 13} 064001.

\bibitem{karimi:14} 
Karimi E, Schulz  S A, De Leon I, Qassim V, Upham J and Boyd R W 2014 \textit{Light: Science \& Applications} {\bf 3}, e167.

\bibitem{cardano:13} 
Cardano F, Karimi E, Marrucci L, de Lisio C and Santamato E  2013 \textit{Opt. Express} {\bf 21} 8815.

\bibitem{beckley:10} 
Beckley A M, Brown T G and Alonso M A 2010 \textit{Opt. Express} {\bf 18} 10777.

\bibitem{karimi:07} 
Karimi E, Zito G, Piccirillo B, Marrucci L and Santamato E 2007 \textit{Opt. Lett.} {\bf 32}, 3053.

\bibitem{cardano:12}
Cardano F, Karimi E, Slussarenko S, Marrucci L, de Lisio C and Santamato E 2012 \textit{Appl. Opt.} {\bf 51}, C1.

\bibitem{anton:01}
Mair A, Vaziri A, Weihs G and Zeilinger A 2001 \textit{Nature (London)} {\bf 412}, 313.

\bibitem{miles:99}
Padgett MJ  and Courtial J 1999 \textit{Opt. Let.} {\bf 24}, 430.

\bibitem{nagali:09b}
Nagali E, Sciarrino F, De Martini F, Karimi E, Piccirillo B, Marrucci L and Santamato E 2009 \textit{Opt. Express} {\bf 17}, 18745.

\bibitem{hammam:14}
Qassim H, Miatto FM, Torres JP, Padgett MJ, Karimi E and Boyd RW 2014 \textit{J. Opt. Soc. Am. B} {\bf 31}, A20.

\bibitem{steve:14}
Roux FS and Zhang Y 2014 \textit{Phys. Rev. A} {\bf 90}, 033835.

\bibitem{gbur:01}
Gbur G, Visser T D and Wolf E 2001 \textit{Phys. Rev. Lett.} {\bf 88}, 013901

\bibitem{amiri:05}
Tavassoly M T, Amiri M, Karimi E and Khalesifard H R 2005 \textit{Opt. Commun.} {\bf 255}, 23.

\bibitem{Tosh:12}
Wakayama T, Komaki K, Otani Y and Yoshizawa T 2012 \textit{Opt. Express} {\bf 20}, 29260.

\bibitem{Tosh:14}
Wakayama T, Rodríguez-Herrera O G, Tyo J S, Otani Y, Yonemura M and Yoshizawa T 2014 \textit{Opt. Express} {\bf 22}, 3306.

\end{thebibliography}
\end{document}